\documentclass[fleqn,usenatbib]{mnras}

\usepackage{newtxtext,newtxmath}

\usepackage[T1]{fontenc}

\DeclareRobustCommand{\VAN}[3]{#2}
\let\VANthebibliography\thebibliography
\def\thebibliography{\DeclareRobustCommand{\VAN}[3]{##3}\VANthebibliography}

\usepackage{graphicx}	
\usepackage{amsmath}	

\title[Low-z GRBs from WDBHs]{White Dwarf-Black Hole Binary Progenitors of Low Redshift Gamma-ray Bursts}

\author[Lloyd-Ronning et al.]{
Nicole M. Lloyd-Ronning,$^{1,2}$\thanks{E-mail: lloyd-ronning@lanl.gov}
Jarrett Johnson,$^{2,3}$
Phoebe Upton Sanderbeck,$^{2,4,6}$
\newauthor \ Makana Silva,$^{1,2,6}$ Roseanne M. Cheng$^{1,5}$ 
\\
$^{1}$Computational Physics and Methods Group, Los Alamos National Lab, Los Alamos, NM 87544\\
$^{2}$Center for Theoretical Astrophysics, Los Alamos National Lab, Los Alamos, NM 87544\\
$^{3}$X Theoretical Design Division, Los Alamos National Lab, Los Alamos, NM 87544\\
$^{4}$Information Systems and Modeling Group, Los Alamos National Lab, Los Alamos, NM 87544\\
$^{5}$University of New Mexico, Albuquerque, NM, 87106\\
$^{6}$Director's Postdoctoral Fellow
}

\date{Accepted XXX. Received YYY; in original form ZZZ}

\pubyear{2024}

\begin{document}
\label{firstpage}
\pagerange{\pageref{firstpage}--\pageref{lastpage}}
\maketitle

\begin{abstract}
 Although there is strong evidence that many long GRBs are associated with the collapse of a massive star, tantalizing results in recent years have upended the direct association of {\em all} long GRBs with massive stars. In particular, kilonova signals in some long GRB light curves as well as a suggested uptick in the rate density of long GRBs at low redshifts (deviating significantly from the star formation rate) suggest that compact object mergers may be a non-negligible fraction of the long GRB population. Here we investigate the contribution of white dwarf-black hole mergers to the long GRB population.  We present evidence for the deviation of the long GRB rate density from the star formation rate at low redshifts, and provide analytic and numerical arguments for why a white dwarf-black hole merger system may be a viable progenitor to explain this deviation. We show the range of parameter space in which the durations, energetics, and rates of these systems can account for a significant sub-population of low-redshift long GRBs. 
 
\end{abstract}

\begin{keywords}
stars: gamma-ray bursts: general -- stars: binaries: general
\end{keywords}



\section{Introduction}

For many decades, we have accumulated observational evidence that has led to a general cognizance of gamma-ray burst (GRB) progenitor systems.  Based on the duration of their prompt gamma-ray emission, energetics, locations in their host galaxies, associated supernovae (or lack there-of), associated kilonovae and gravitational wave signals (or lack there-of), the following paradigm has emerged: long gamma-ray bursts (lGRBs) -  GRBs with prompt gamma-ray emission lasting longer than two seconds -  appear to be associated with the death of massive stars \citep[e.g.][]{Gal98,Hjorth03,WB06,HB12}. Meanwhile short GRBs (sGRBs) - GRBs with duration lasting less than about two seconds) - are associated (at least in some cases) with the collision of two neutron stars or a neutron star and a black hole \citep{Berg14, fong15, Ab17}. \\

However, a number of recent results have called into question this general picture, and there still remain many open questions regarding the nature of both lGRB and sGRB progenitor systems. For example, recent observations suggest that some long GRBs may be a result of a double neutron star merger based on a potential kilonova signal in their lightcurves, opening up the possibility that GRBs with prompt gamma-ray emission lasting 10's of seconds may indeed be produced by compact object binaries \citep{Yang22,Rast22,Tro22,Gott23,Lev23,Yang24}.  Additionally, recent studies using machine learning techniques to explore the classification of GRB progenitors \citep{Zhu24, Dim24} have found that there may be be significant ``contamination'' or overlap between these two (i.e. short and long GRB) samples \citep[this overlap had been pointed out previously, although in the context of suggesting a third class of GRBs; see, e.g.][]{Horvath16}.  \\

Interestingly and potentially related to this issue, several studies have seen an uptick in the rate density of long GRBs at low redshifts \citep{Pet15, Yu15, Tsv17, LM17, LR19, Le20, LR20, Has24, Pet24}.  Recently, \cite{Li24} examined a subset of GRBs with extended emission and found they can be separated into two ``classes'' or groups according to the slope and intercept of their $E_{iso}-E_{p}$ relation, where $E_{iso}$ is the isotropic emitted energy and $E_{p}$ is the peak of the prompt gamma-ray spectrum \citep[see also][]{Singh2024}.  Given their classifications, they find that the rate density of one sub-sample (which they name EE2) shows a strong increase at low redshifts while the other (sub-sample EE1) appears to more closely track the star formation rate (decreasing at low  redshift). They interpret this as evidence of the presence of a distinctly different progenitor at low redshifts (relative to the progenitor responsible for the EE1 sub-sample). Others have suggested that the low redshift uptick may be due to the contribution from a compact object merger progenitor at these redshifts \citep{Pet24}, distinct from a system that more closely follows the star formation rate (e.g. a massive star).  As such, it is important to consider the unique contribution of these compact object merger populations to the long GRB rate density.\\


There are a number of potential compact object merger systems that could contribute to the low redshift uptick of the lGRB rate density, including double neutron star (DNS) mergers, neutron star-black hole (NSBH) mergers, double white dwarf (WD) binary mergers, white dwarf-neutron star (WDNS) mergers, and white dwarf-black hole (WDBH) mergers.  We have briefly discussed above the potential contribution from DNS and NSBH binary mergers; those systems remain viable contributors to the low redshift uptick, although it is not easy to create the long-lived prompt gamma-ray emission in these scenarios due to the lack of available material in the disk around the central engine \citep[e.g.][]{JP08}\footnote{This is under the assumption of a magnetically powered jet launched from a black hole-disk central engine.}.  Double WD binary mergers may be too weak, without enough material/fuel, to power a long GRB \citep[see, e.g.,][]{Krem21,YT22}.  A WDNS binary merger has been suggested as a potential progenitor for at least one GRB \citep{Zhong23}, as long as the WD component has a mass $> 1M_{\odot}$ \citep[on the other hand, ][has shown that for lower mass WDs the systems can survive and become ultracompact X-ray binaries]{Bob17}. The rates and delay-time distributions of WDNS binaries are explored in detail in \cite{Toon18}, who show these systems typically merge in $< 1$ Gyr and have a wide range of offsets from their host galaxies.\footnote{We note that $1$ Gyr is shorter than the time between the peak of the star formation rate at $z \approx 2$ and the apparent uptick at $z <1 $.}  In principle these systems may contribute to the low redshift uptick of long GRBs, although rate estimates are uncertain.\\

In this paper we explore how white dwarf-black hole (WDBH) binary mergers contribute to the low redshift uptick in the rate density of lGRBs.  These models have been suggested as progenitors for lGRBs in a few studies \citep{Fry99, Dong18}, but their population synthesis and contribution to the low redshift population of long gamma-ray bursts has not yet been fully scrutinized.  We investigate the possibility that white dwarf black hole mergers have the energetics, time scales, and overall necessary rates to accommodate the uptick in the low redsfhift rate density of GRBs.\\

 Our paper is organized as follows.  In \S 2 we recap the evidence that suggests a potential distinct population of low-redshift long GRBs, including an excess in the rate density at low redhifts. In \S 3, we summarize previous works and present order of magnitude calculations that show WDBH binaries have the necessary energetics and timescales to power a long GRB.  We also discuss the many factors that can play a role in their merger timescales. In \S 4, we present results of population synthesis simulations which suggest that the rates of these systems per galaxy over a range of metallicities align with observational constraints.  We provide analytic calculations of the WDBH rate density, using empirical estimates of delay-time distributions of compact object mergers, and present the range of parameter space for which WDBH mergers can explain the low redshift uptick in the observed rate density of long GRBs. Finally we present a summary and our main conclusions in \S 5.

 \begin{figure*}
	\includegraphics[width=15cm]{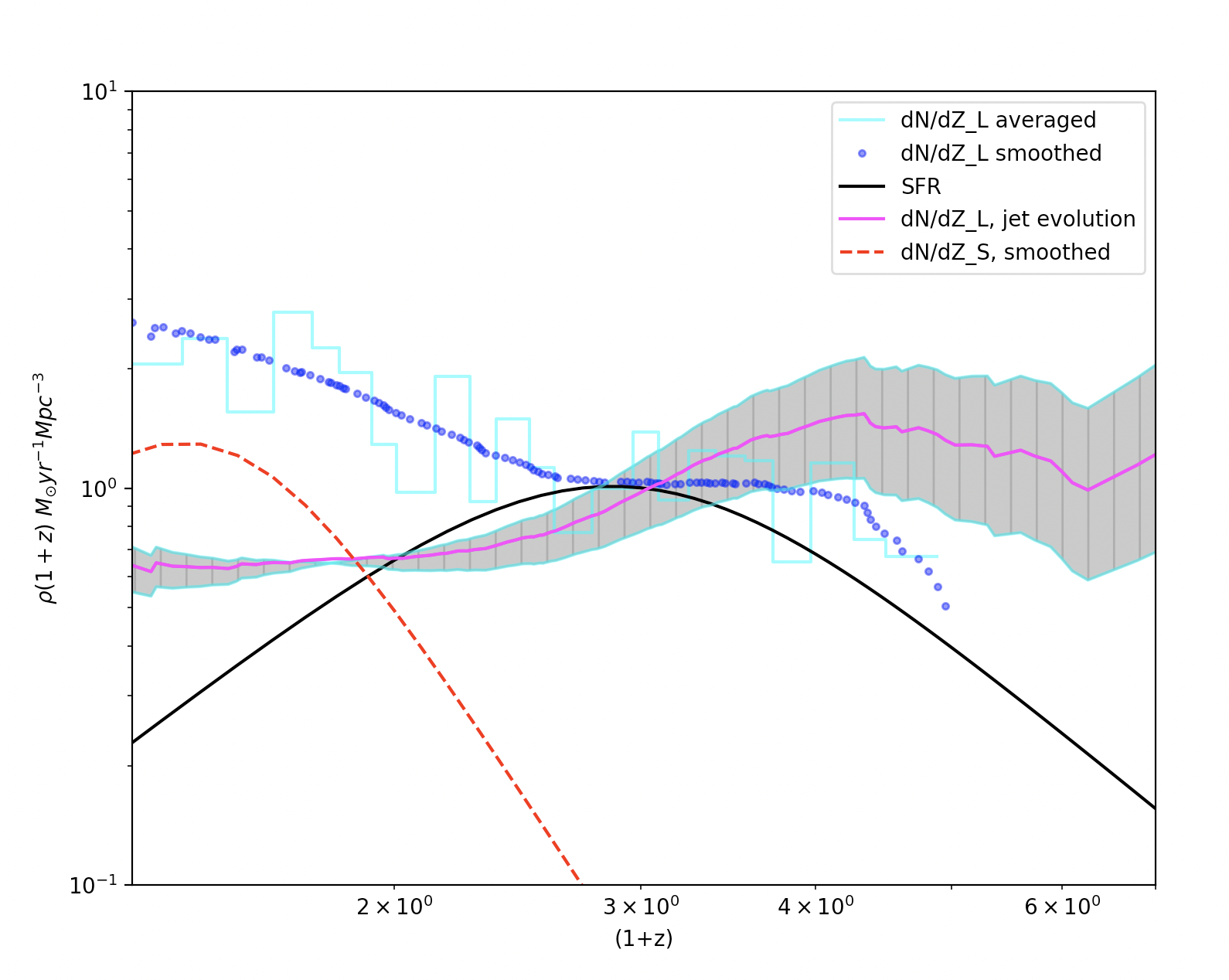}
    \caption{Long (L) GRB star formation rate density without accounting for jet opening angle evolution (cyan and blue curves), accounting for jet opening angle evolution (magenta and gray regions), and short (S) GRBs rate density (red dashed curve).  The black solid line indicates the star formation rate according to \protect\cite{MD14}.  The long GRB rate densities are normalized to the peak of the star formation rate at $(1+z)$ = 3, while the short GRB rate density is normalized at a redshift of $(1+z) = 2$.  As noted in \protect\cite{LR20} the uptick at low redshifts is somewhat reduced if beaming angle evolution is accounted for.}
    \label{fig:grbrate1}
\end{figure*}

 \section{Data}
 
 Figure~\ref{fig:grbrate1} shows the estimated or predicted star formation rate density from the observed GRB rate, relative to the global star formation rate (SFR) as determined by \cite{MD14} (black line). 
 
  \begin{equation}
     \dot{\rho}_{\rm SFR\_GRBs}(z) = (\dot{dN}/dz)({\rm f_{beam}(z)})  \left(\frac{(1+z)}{dV/dz} \right) \frac{1}{\epsilon(z)}
  \end{equation}   \\
 
 \noindent where $\dot{dN}/dz$ is the GRB rate (accounting for the GRB luminosity evolution and detector trigger selection effects), $\epsilon(z)$ parameterizes the fraction of stars that make GRBs (and in principle can evolve with redshift), and $f_{\rm beam}(z)$ is a factor ($>1$) that accounts for the gamma-ray burst beaming angle \citep[and its evolution with cosmic redshift; e.g.][]{LR19}.  The factor $dV/dz$ is the cosmological volume element given by:
 \begin{equation}
\begin{split}
    dV/dz = & 4 \pi (\frac{c}{H_{o}})^{3} \bigg[\int_{1}^{1+z} \frac{d(1+z)}{\sqrt{\Omega_{\Lambda} + \Omega_{m}(1+z)^{3}}}\bigg]^{2} \\
    & \times \frac{1}{\sqrt{\Omega_{\Lambda} + \Omega_{m}(1+z)^{3}}}
\end{split}
\end{equation}
\noindent Removing the $1/\epsilon (z)$ factor, one can think of $\dot{\rho}_{\rm SFR\_GRBs}(z)$ as the rate density of GRB progenitor formation.\\

 The cyan and blue lines show the rate density derived from observations of long GRBs, but  without correcting for jet beaming angle evolution.  The magenta and gray hashed region show the rate density, accounting for the fact the high redshift GRBs may be more narrowly collimated than low redshift GRBs \citep{LR20}; because in general we only observe the GRBs whose jets are pointed in our direction this means we are missing a higher fraction of GRBs in the high redshift universe.  The red dashed line is the rate density derived from observations of short GRBs (see also \cite{Dai21} for a calculation of the rate density of short GRBs using non-parameteric methods).  The long GRB curves are normalized to the \cite{MD14} star formation rate at its peak, while the short GRB curve is normalized to the star formation rate at a redshift of $(1+z) \sim 2$; this is an arbitrary choice, but attempts to account for the delay time between star formation and merger time in a double neutron star merger model for short GRBs  \citep{Nak06,Berg07,HY13,WP15,Ana18,Zev22}.\\

 Although accounting for the jet beaming angle evolution decreases the extent of the low redshift uptick, it is still present in all cases for the long GRBs.  The short GRB rate density, as expected, peaks at lower redshifts (most likely due to the time it takes the stars in this progenitor model to evolve and then merge). Again, the shape of the short GRB rate density seems to suggest that there might exist a contribution from compact object mergers to the low redshift uptick in the long GRB population. \\

 We note that, in this figure the underlying GRB rate density - accounting for detector selection effects and luminosity evolution - was determined using the non-parametric statistical methods of \cite{LB71, EP92, ep98}.  Others \citep{WP10, Lien14, Pesc16} which have used parametric methods (which often makes an implicit assumption that the GRB rate follows the star formation rate more closely) see a smaller or no uptick at low redshifts\footnote{We note, however, that \cite{Lan19} also used non-parametric methods to estimate the GRB rate density and luminosity function (including evolution) and found only a mild uptick at low redshifts, although their luminosity evolution agrees well with \cite{LR19}.}. Additionally, we note that if the formation efficiency of GRBs is drastically reduced at low redshifts (i.e. on average higher metallicities), this can also reduce the severity of the uptick and align the GRB rate density more closely with the star formation rate density \citep{Per16}. Finally, \cite{Lu24} suggest that non-parametric methods can have degeneracies in the functional form of the underlying correlation so caution must be taken in drawing too strong of conclusions on the exact form of that function. \\

Given the possibility of a distinct population of low redshift lGRBs, we have examined the isotropic energy distribution and duration of the prompt gamma-ray emission (corrected for time dilation) for the subset of bursts with  $(1+z) < 2$.  This cutoff is chosen simply because that is the redshift below which the deviation from the star formation rate begins to appear, though it does not necessarily delineate any GRB progenitor sub-populations in a robust way.  We find the average value of $E_{iso}$ is lower for this subset of low redshift bursts relative to the whole population at a marginally statistically significant level ($2.5 \sigma$).  There is no statistically significant difference in the values of the durations between these samples.  Again, this is not necessarily surprising because we have not attempted to delineate in any way the sub-sample of low redshift lGRBs that maybe be coming from WDBH mergers and so there would be mixing between such a sub-sample and ``traditional'' lGRBs that come from the collapse of massive stars. \\

As mentioned in the Introduction, \cite{Li24} examined a population of GRBs with extended emission and found that this population could be further classified into two groups (which they named ``EE1'' and ``EE2'') based on the slope of the correlation between isotropic equivalent energy $E_{iso}$ and the spectral peak energy $E_{p}$.   Their EE2 sub-sample shows a rate density that is significantly enhanced at low redshifts suggesting that perhaps this sub-sample originates from a distinct progenitor more prevalent at low redshifts. \\

\begin{figure}
	\includegraphics[width=\columnwidth]{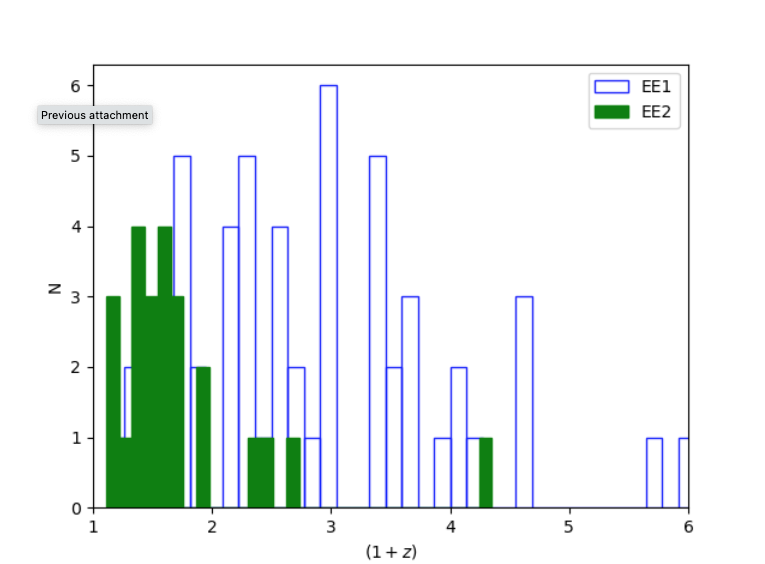}
    \caption{Observed redshift distribution of EE1 vs EE2 samples, as defined by \protect\cite{Li24}. They suggest the EE2 sample may be coming from a distinct progenitor system compared to the EE1 sample.}
    \label{fig:EE1EE2redshift}
\end{figure}

\begin{figure*}
	\includegraphics[width=\columnwidth]{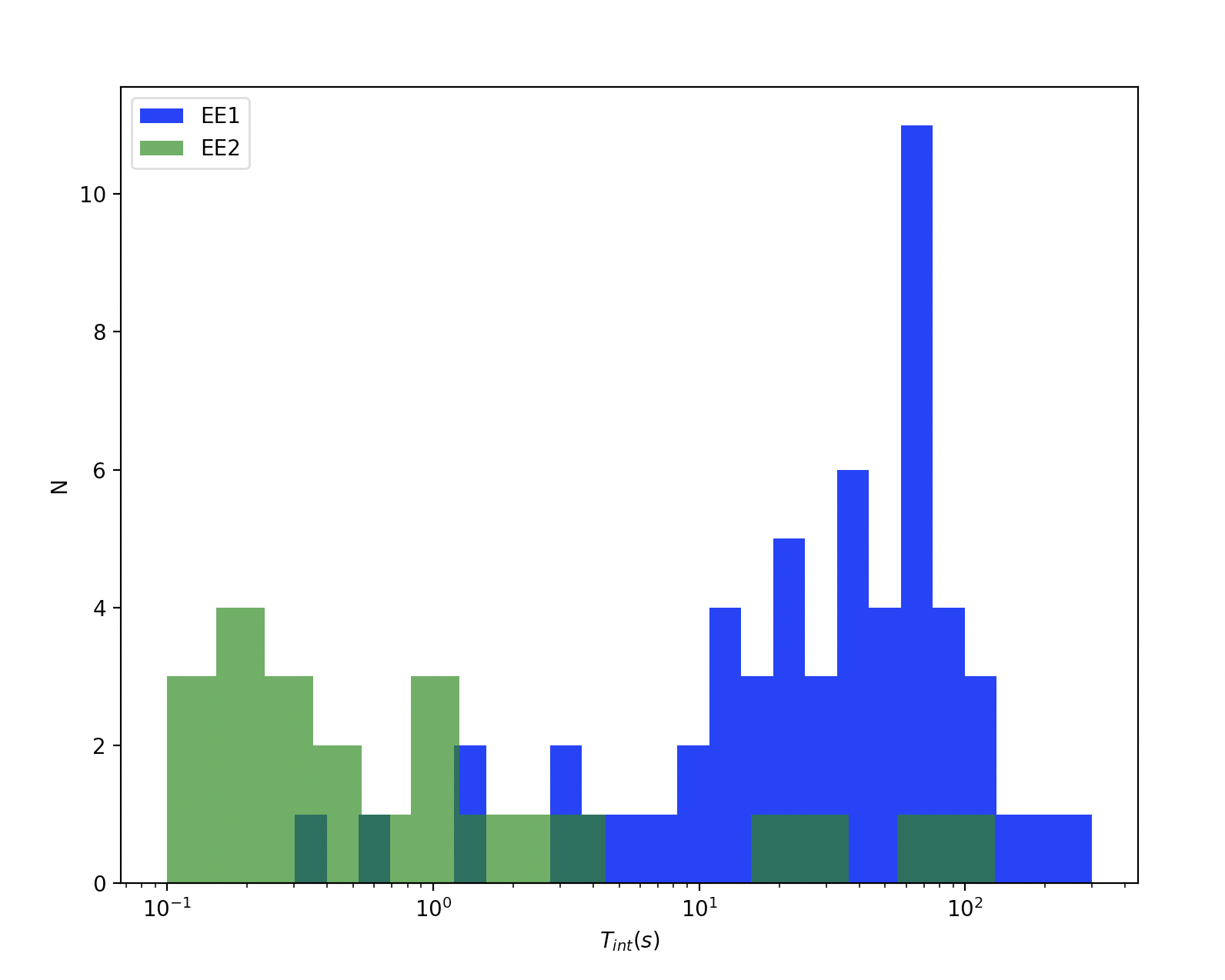}\includegraphics[width=\columnwidth]{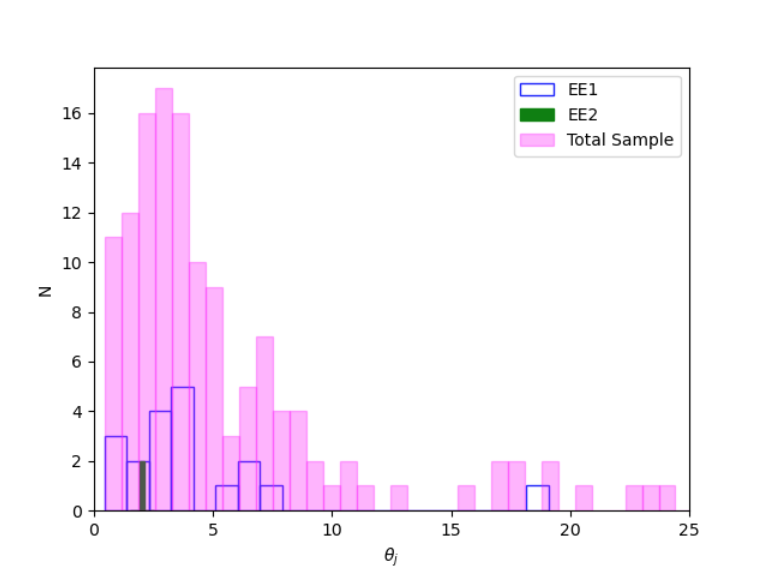}
    \caption{ {\bf Left panel:} Plot of intrinsic gamma-ray duration for the so-called EE1 (blue histogram) and EE2 (green histogram) samples from \protect\cite{Li24}.  The EE2 sample has a significantly shorter duration than the EE1 sample. {\bf Right panel:} Plot of the jet opening angle for the EE1 (blue histogram) and EE2 samples (green histogram), as well as all lGRBs for which this data exist (pink histogram). 
 Note that the EE2 sample has only a single jet opening angle measurement. }
    \label{fig:EE1EE2dur}
\end{figure*}

We examined certain characteristics of the \cite{Li24} EE1 and EE2 sub-samples when the data were available, where - again - the former sample appears to follow the star formation rate while the latter exhibits an uptick at low redshifts relative to the star formation rate.  We show the observed redshift distributions of their EE1 and EE2 samples in Figure~\ref{fig:EE1EE2redshift}.  We see that GRBs in their EE2 sample have significantly shorter prompt gamma-ray duration, shown in the left panel of Figure~\ref{fig:EE1EE2dur}.  We also looked at the jet opening angles in their sub-samples.  There are 19 GRBs in their EE1 sample that have jet opening angle measurements, with a distribution that is not significantly different from the global long GRB population (Figure~\ref{fig:EE1EE2dur}, right panel).  That only 2 GRBs in their EE2 sample had jet opening angle measurements could suggest a potentially much wider jet for this sample.  However, many observational selection effects come into play when measuring jet opening angle and we simply do not have the numbers to test this hypothesis rigorously.\\

Based on studies that have suggested a dichotomy in progenitors between long GRBs with and without radio afterglows \citep{LR17,LR19rnr, Chak23}, we also looked at the presence or absence of radio emission in the EE1 and EE2 samples.  Again, the numbers are small but we found that of those GRBs in the \cite{Li24} samples that have a radio afterglow, 11 are from their EE1 sub-population and only 1 is from the EE2 population.  This is consistent with the conjecture that radio bright GRBs are the result of the collapse of a massive star in a dense gaseous environment (star forming regions) while radio dark GRBs are in more tenous environments as one might expect in a WDBH merger progenitor scenario.\\


 If a low redshift sub-population of lGRBs might have a significant contribution from compact object binary merger progenitors, it is valuable to consider the host-galaxy offsets of low redshift lGRBs, as compact object binary systems may receive a significant kick upon formation of the neutron star or black hole component (from the supernova that made it) and therefore tend to migrate well outside their host galaxies by the time they merge (and create the GRB).  Meanwhile, massive star collapse progenitors tend to be found in the star forming regions of their host galaxies, with relatively small offsets \citep{Bloom02,Ly17}.  We might expect that if there exists a low-redshift population of lGRBs from compact object mergers, the host galaxy offsets of low redshift GRBs may be on average slightly larger than those of high redshift GRBs.  
 
 However, according to \cite{BBF16} there is no clear trend of host-galaxy offset with redshift for a sample of 100 or so lGRBs (see their Figure 4).  When looking at the offsets of the \cite{Li24} EE1 and EE2 samples, it is hard to draw any conclusions: The EE1 sample has only 6 offset measurements but having a look at the so-called EE1 and EE2 samples, there were 6 GRBs in the EE1 sample with offset measurements and with values ${0.7, 0.81, 0.82, 1.73, 3.42, 5.20}$ kpc for the physical offset and ${0.27, 0.3, 0.87, 1.09, 1.6, 3.425}$ for the normalized offset (normalized to the half-light ratio of the galaxy). These numbers appear to align with what is observed in lGRBs where most offsets appear to be within less than a few kpc or roughly within the half-light radii of their host galaxies \citep{Bloom02}, as opposed to sGRBs, many of which have observed offsets that are 10's of kpc from the center of their host galaxies \citep{Fong22}. There is only one GRB in the EE2 sample that had an offset measurement, with values $0.992$ kpc for the physical offset and $0.765$ for the normalized offset.  Hence, there is nothing in the sparse data to suggest any unusual trends with host-galaxy offset, although of course the small numbers make it difficult to draw any definitive conclusions about the existence or presence of a trend. We note that it is the supernova process with significant ejecta mass in addition to heavy element production that is believed to be responsible for large natal kick velocities in neutron stars \citep{Wong13, Bray16}, Therefore, in general black holes may have smaller natal kick velocities than neutron stars (so that a large host galaxy offset may not be observed), although recent observations of mis-aligned spins from gravitational wave observations of black hole binaries have called this view into question \citep{Osh17}.  \\ 

 Putting all of this together, we suggest that a WDBH binary system may have the characteristics of both long and short GRBs, and be able to explain the trends (or lack there-of) in the data described above.  In the next sections we explore analytically and with population synthesis models the plausibility of a WDBH merger progenitor for lGRBs, and as a contributor to the low redshift uptick in the lGRB rate density.

\section{Model}

\subsection{Energetics and Timescales}

In a WDBH merger model, the relativistic jet that produces GRB emission is launched from a black hole-accretion disk central engine that is formed when the white dwarf overflows its Roche lobe and is tidally disrupted by the black hole, with its stellar material circularizing in a disk around the stellar mass black hole \citep[for the physics of tidal disruption of a WD around massive black holes, see, e.g.][]{Mac14, LX21}.  \\

\cite{Fry99} provide analytic estimates, along with numerical SPH simulations, of the amount of material in the disk, the accretion rate, and the angular momentum in the disk.  They find \cite[using the equations of][]{Egg83} that the white dwarf is tidally disrupted when it reaches an orbital separation from the black hole between about $3$ -- $5$ $\times 10^{9}$ cm, with nearly all of the mass of the white dwarf circularizing (roughly about 1 M$_{\odot}$ of material) and an accretion  rate of 0.05 M$_{\odot}$  s$^{-1}$.  \\

Observations of GRBs require that the isotropic equivalent energies, $E_{iso}$, lie in the range of $10^{50}$ to $10^{54}$ ergs. In reality, the energy released $E_{\gamma}$ is concentrated in a relatively narrow jet and is closer to  $\lesssim 10^{50}$ erg. In other words, $E_{\gamma} = E_{iso}(1-{\rm cos}(\theta_{j}))/2$, where $\theta_{j}$ is the opening angle of the jet. From an energy conservation standpoint alone, the orbital energy of the binary, the rotational energy of the black hole, and/or the mass energy of the white dwarf are all sufficient energy reservoirs to account for the GRB energy budget.  However, we need to consider in more detail how this energy is extracted from the system.  A viable mechanism to is to have the WD material tidally disrupted and its material circularized in a jet around the black hole; magnetic fields in the disk will rapidly grow through the magneto-rotational instability \citep{V59, Chan60,AH73,BH91}, and wind up (through frame-dragging) along the spin axis of the black hole. This can lead to a magnetically launched relativistic jet, which ultimately extracts the spin energy of the black hole through the so-called Blandford-Znajek (BZ) mechanism \citep{BZ77,MT82}.  The luminosity of the jet is given by:
\begin{equation}
\centering
\begin{aligned}
L_{BZ} & =  (kfc^{5}/64\pi G^{2})a^{2}\phi^{2}M_{BH}^{-2}  
\end{aligned}
\end{equation} 
\noindent where $k$ is a geometrical factor related to the magnetic field geometry (of order $\sim 0.05$), $f$ is a factor of order unity, $c$ is the speed of light, $\phi$ is the magnetic flux on the black hole, $a = Jc/GM_{BH}^{2}$ is the black hole spin parameter, where $J$ is the black hole angular momentum, and $M_{BH}$ is the black hole mass $M_{BH}$. 

The jet power depends on the spin of the black hole, the magnetic flux, and the mass of the black hole, although we note the complicated interplay between all three of these variables.  We can simplistically write down, under the assumption of flux conservation, the magnetic flux in terms of magnetic field strength $B$ and mass of the black hole\footnote{ $\phi \approx 4 \pi B R^{2}$, where $B$ is the magnetic field, and $R$ is the Kerr radius given by $R = GM/c^{2} + \sqrt{(GM/c^{2})^{2} - a^{2}}$}, which leads us to the following expression for the observed jet luminosity \citep{mck05, TNM10, Tch15, LR19bz}:
 \begin{equation}
\centering
\begin{aligned}
& L_{GRB} \approx  \\ 
 & 10^{50} {\rm erg} (\eta/0.1) (a/0.9)^{2}(B/10^{16}{\rm G})^{2}(M_{BH}/5M_{\odot})^{2}
\end{aligned}
\end{equation} 
  
\noindent and where we have used an efficiency factor $\eta$ between the BZ jet power and the observed GRB jet luminosity, $L_{GRB} = \eta L_{BZ}$.  Provided the disk can sustain the necessary magnetic flux\footnote{Many studies \citep{Ros02, PR06, Ob10, tch11, ZM13, Lis18,LTQ18, TAM18} have shown that indeed the disk can sustain a very high magnetic flux.} and the black hole spin is sufficient (with the spin parameter $\gtrsim 0.5$. Our population synthesis results described below show that the vast majority $>90 \%$ of the WDBH systems we simulated have a black hole spin parameter > 0.8 (see the bottom right panel of Figure 4), where we have used a prescription for the black hole spin based on the core mass as described in \cite{Bel20}.\\

We also require that the jet is sustained long enough to explain the duration of the prompt gamma-ray burst itself.  The jet lifetime - which is directly correlated with the duration of the GRB prompt emission - is roughly proportional to the mass of the disk divided by the accretion rate:
\begin{equation}
    T_{{\rm jet}} \approx M/\dot{M}
\end{equation}

Given that we have approximately a solar mass of material and using the estimate from \cite{Fry99} of an accretion rate of about $5 \times 10^{-2} M_{\odot}/s$, this will allow for a roughly $20 s$ GRB, well within the ``long'' class of GRBs (duration $>2s$).  The accretion rate in this model is an uncertainty; for GRBs in general, studies have shown that accretion rates can span a wide range of values \citep{Pop99, KNJ08, Tch15, JJN22, JJ22}, from as low as $10^{-3}  M_{\odot}/s$\footnote{This is particularly true in the Magnetically Arrested Disk (MAD) model in which magnetic flux built at the horizon halts the accretion onto the black hole; the jet is launched by the magnetic pressure which is sustained by the rotational energy of the black hole \citep{Tch15}.}  (which will produce longer GRBs) to as high as $10  M_{\odot}/s$ (which will produce shorter GRBs, for a given amount of material in the disk).  In the WDBH model, again, we only have about $1 M_{\odot}$ of material available (although this number can range from about $0.2$ to $1.4 M_{\odot}$), but - given the large range of possible accretion rates - this still may accommodate GRBs ranging from sub-second up to hundreds (even thousands) of seconds.   Hence, we suggest that this model can reasonably accommodate the durations of lGRBs.

  
\subsection{Merger Timescale}

A key factor when attempting to account for the apparent uptick of the low redshift GRB rate density from a compact object merger progenitor is properly factoring in the delay time from the formation of the binary until the merger itself (when the GRB is produced).  In other words, compact object merger populations - including WDBH mergers - are expected to follow the star formation rate {\em convolved with a delay time distribution}.  \\

 However, this distribution is highly uncertain from a theoretical standpoint.  It is a combination of the stellar evolution timescale (i.e. the time for the system to form and evolve from the main sequence lifetimes of the stars in the binary), and the orbital decay timescale until merger.  This latter timescale depends on what is causing the loss of orbital angular momentum in the system, which can happen through multiple channels.  For example, gravitational wave emission (which depends on the masses of the binary components, their separation, and eccentricity) as well as losses due to gas and dynamical friction (which depends on a a number of complicated environmental factors) can all contribute significantly to this timescale\footnote{For reference, we note the Keplerian time $t_{\rm orb} = (4 \pi r^{3}/GM)^{1/2}$ is very short ($< 1$ yr) for separations $r$ on the order 10's of solar radii and for a total mass $M$ of the system of $10 M_{\odot}$.}. \\

\noindent We define the merger timescale as follows:
\begin{equation}
    t_{{\rm merge}} =  t_{{\rm evolve}} + {\rm min}(t_{{\rm loss}})
\end{equation}
\noindent where $t_{\rm evolve}$ is the timescale for the WDBH system to form, and ${\rm min}(t_{{\rm loss}})$  is the shortest timescale over which orbital angular momentum is lost up until the merger.  The evolution times of these systems, as determined by our population synthesis simulations discussed in \S4 below, are in general around a few hundred Myr. This is the timescale, if the initial conditions permit, for a main sequence binary system to form a black hole and white dwarf component, with the timescale dominated by the stellar evolution of the lower mass star (i.e. the star that becomes the WD). However, we note that this is for the high end of the mass range of white dwarfs ($> 1 M_{\odot}$; this is to allow for sufficient material in the disk that forms during the merger and aligns with our estimates in \S 3).  Lower mass stars will take much longer ($\sim 10$ Gyr) to evolve and can provide a significant contribution to the WDBH population (and the delay time). The orbital angular momentum loss timescales can span a wide range of values, and there are many different channels that contribute to this loss timescale, including gravitational wave emission, common envelope evolution, mass loss, dynamical friction, gas dynamical friction, and more.  We discuss the contributions from a few of these channels below. \\


\noindent {\bf Gravitational Wave Loss:}\\
The loss timescale due to gravitational wave emission is given by:
    \begin{equation}
        t_{GW} \approx \frac{E_{orb}}{dE/dt_{GW}}
    \end{equation}
where $E_{orb}$ is the orbital energy
\begin{equation}
    E_{orb} \approx G M_{WD} M_{BH}/r
\end{equation}

\noindent and 

    \begin{equation}
   dE/dt_{GW} = \frac{32 G^{4}(M_{WD}+M_{BH})M_{WD}^{2}M_{BH}^{2}}{5c^{5}r^{5}} f(e)
\end{equation}

\noindent where $M_{WD}$ is the mass of the white dwarf, $M_{BH}$ is the mass of the black hole, $r$ is the separation, and $f(e)$ is a factor that depends on eccentricity which is of order 1.  Combining equations 8 and 9, we find:
\begin{equation}
    t_{GW} \approx \frac{5c^{5}r^{4}}{32f(e)G^{3}(M_{WD}+M_{BH})M_{WD}M_{BH}}
\end{equation}

This timescale is highly dependent on separation.  For a binary separation of 10 $R_{\odot}$, a 10 $M_{\odot}$ mass black hole, and 1 $M_{\odot}$ white dwarf, this merger time due to gravitational wave radiation is $\approx 10^{11}$ year (i.e. longer than the age of the universe). On the other hand, for a separation of a $0.1 R_{\odot}$, this timescale can be as short as about $10^{3}$ year.  We have found in our population synthesis results (discussed below) that a significant fraction of these binaries tend to form at small $< R_{\odot}$ separations, although most of the population has separations $\ge R_{\odot}$.  Therefore, given this very large timescale to lose energy from gravitational wave emission for much of the WDBH population, it is crucial to consider other mechanisms that may lead the orbit to decay more quickly and merge within a Hubble time.  And there are indeed a number of physical processes expected to contribute to this frictional loss timescale. \\

 For example, common envelope evolution is a widely accepted way of drastically reducing the orbital separation of a binary system and bringing a compact object binary merger time to within a Hubble time \cite[e.g.,][we discuss this further below when we invoke a delay time distribution for our WDBH mergers]{Dom12, Mac18}.  Binaries embedded in a gaseous or densely populated environment can also lead to a drastic loss of orbital angular momentum and cause the binary system to merge well within a Hubble time \citep{Zahn08, Roz22, Roz23}.   The precise timescale of these processes is system- and environment-dependent, and therefore can span a a very large (and unconstrained) range, which we discuss briefly now.  \\

\begin{figure*}
	\includegraphics[width=\columnwidth]{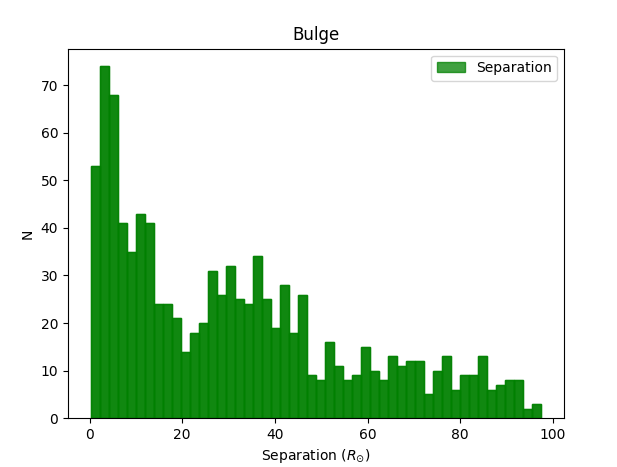}  \includegraphics[width=\columnwidth]{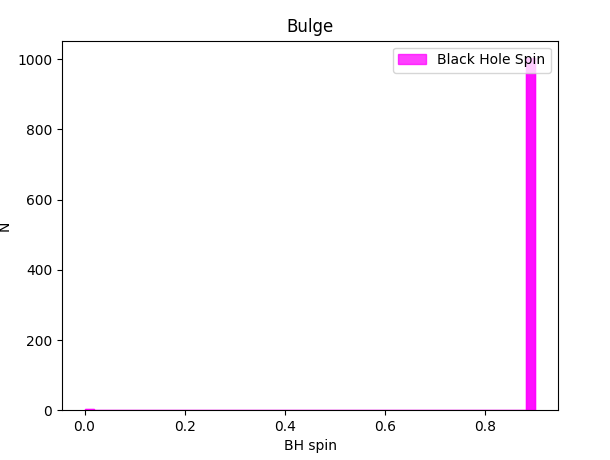} 
    \caption{Distributions of WDBH binary separations (left panel) and spin of the black hole component (right panel),  for a population synthesis simulation at metallicity $Z=10^{-2}$.  The y-axis has arbitrary normalization.}
    \label{fig:example_figure}
\end{figure*}

\begin{figure}
	\includegraphics[width=\columnwidth]{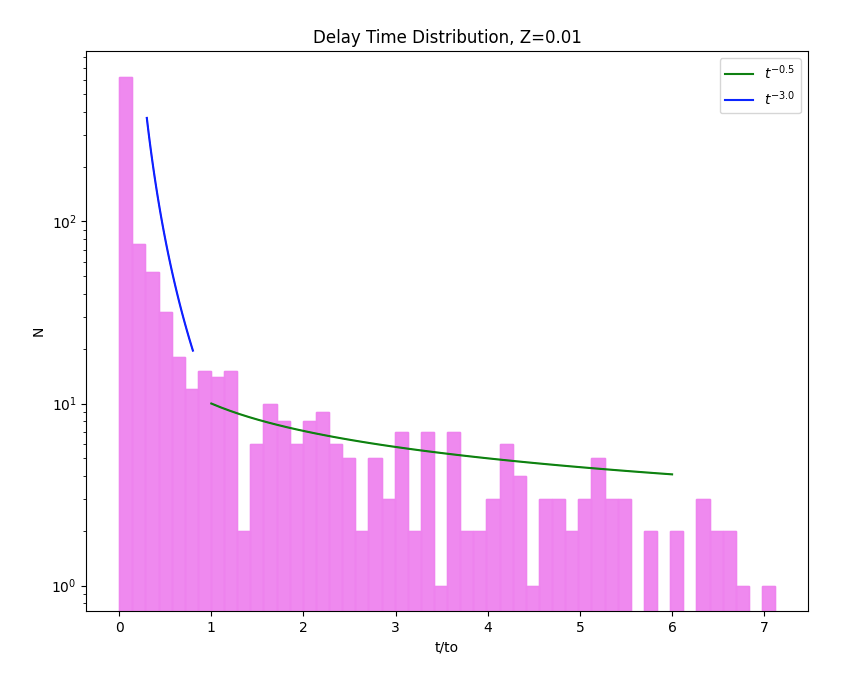}    
    \caption{Delay time distribution (DTD) for a population synthesis simulation at metallicity $Z=10^{-2}$.  The y-axis has arbitrary normalization.  The blue line shows a DTD with a power law index of $-3$, while the green line shows a power-law index of $-0.5$.}
    \label{fig:example_figure}
\end{figure}

\noindent {\bf Dynamical Friction:}\\
 Stars moving through a cluster with velocity ${\rm v}$ will experience deceleration in the direction of their motion from the gravitational forces of the other stars \citep{Chan43}. The timescale is the relaxation time of the system which is related to the total number of stars in a cluster $N$ and the crossing time $t_{cr} = R_{sys}/{\rm v}$, where $R_{sys}$ is the size of the system, and is given by:
 \begin{equation}
     \frac{N}{6ln(N)}t_{cr}
 \end{equation}
 
 For a velocity of $\sim 10^{6}cm/s$, a typical size of cluster $10^{20}cm$, and $N \approx 1000$, this timescale is 10's of Myr. \\ 
  
\noindent {\bf Gas Dynamical Friction:}\\
 \cite{Ost99} computed the gas dynamical friction force on a body with mass $M$ traveling at a velocity ${\rm v}$ through a uniform medium with mass density $\rho$.  This drag is due to the gravitational force between the body and its wake, and can be expressed as:
\begin{equation}
    F = -l \times 4\pi \rho_{\rm gas}(GM)^{2} {\rm v}^{-2}
\end{equation}
\noindent where $l$ is an expression related to the Mach number $M_{cs}$, and has limiting values $l \rightarrow M_{cs}^{3}/3$ for $M_{cs} <<1$, and $l \rightarrow ln({\rm v} t/r_{\rm min})$ for $M_{cs} >> 1$. They found that this drag force is more efficient than the \cite{Chan43} formula for dynamical friction in a collisionless medium as long as $M_{cs} > 1$. \\



We note that \cite{Suzu24} has considered gas dynamical friction in a non-linear regime and find that in the sub-sonic gas, the gas dynamical friction can be even higher than what is estimated by linear theory while at higher Mach number, their results agree closely with linear theory.  This means that we may get an even more rapid decay of the orbit than expected from the \cite{Ost99} results when in the sub-sonic regime.\\

\noindent {\bf Magnetic Braking:}\\
Magnetic braking can also play a significant role in the loss of angular momentum of a binary system \citep{VZ81}.  The system loses angular momentum due to the loss of mass from a magnetically coupled stellar wind, and significant decrease in the orbital angular momentum can happen on a timescale of about 1 Gyr.\\

\noindent {\bf Other Loss Processes:}\\
There are some suggestions that dark matter density spikes exist around stellar mass black holes and can explain rapid orbital decay times \citep{Chan23, Ire24, 2024NuPhB100316487B}. Specifically,  \cite{2020PhRvD.102h3006K} show that a binary system evolving in various dark matter environments \citep[dark matter spikes, cusps, mounds;][]{2024NuPhB100316487B} would experience dynamical frictional similar to gaseous environments, leading to shorter orbital decay times. If one also models dark matter as an ultra light boson, then the dark matter can form quasi-bound states around black holes \citep[``gravitational atoms,''][]{2019PhRvD..99d4001B}. This opens the possibility for energy loss (and thus increased orbital decay rates) via ``ionization'' of such bound states as the energy from a binary system is extracted to excite the lowest energy levels.
\\

\noindent {\bf Empirical Estimates:}\\
 Given the numerous possibilities and vast parameter space (which is relatively unconstrained from first principles) for potential loss of angular momentum in a binary system, we need to consider other ways to constrain the merger timescale.  One avenue to do so is to use the estimated delay time distributions (DTD), inferred from observations of compact object binary mergers.  A number of studies have attempted to constrain this distribution for neutron star mergers under the assumption that they are the progenitors of short GRBs \citep{Nak06, Berg07, JL10, HY13, WP15, Ana18, Bel18, Chr18, Broe22, Santo22}. For example, \cite{Zev22}, using observations of short GRBs, infer a power law for the delay time distribution:
 \begin{equation}
     p(t_{d}) \propto t^{-\beta} \ \  t_{\rm min} < t < t_{\rm max}
 \end{equation}
 \noindent where $t_{d}$ is the delay time to merger.  They find a $\beta = 1.83^{+0.35}_{-0.39}$, $t_{min} = 184$ Myr, and $t_{max} \gtrsim 7.4$ Gyr.  They suggest that common envelope evolution and mass transfer in the binary are the primary physical reasons for the steepness of the DTD.  Similarly, \cite{Maoz24} constrained the DTD of NS binary systems using observations of recycled millisecond pulsars.  They find that their sample can be divided into two populations: a so-called fast population in which the DTD $\propto t^{-1.9}$ and a slow population with the DTD $\propto t^{-1.1}$, with the former having an an exponential cutoff below $t \sim 300$ Myr.  \cite{FM20} used observations of TypeIa supernovae to constrain the DTD of double WD progenitor systems.  Their best fit to the a power-law parameterization is also roughly $t^{-1}$. \\

 In what follows, we use these empirical estimates of DTDs as a general guide to parameterize the DTD of WDBH binary mergers (although we allow for a wider range of parameterizations of the DTD), and subsequently estimate the WDBH rate density as a function of redshift.

\begin{figure*}
\includegraphics[width=8cm]{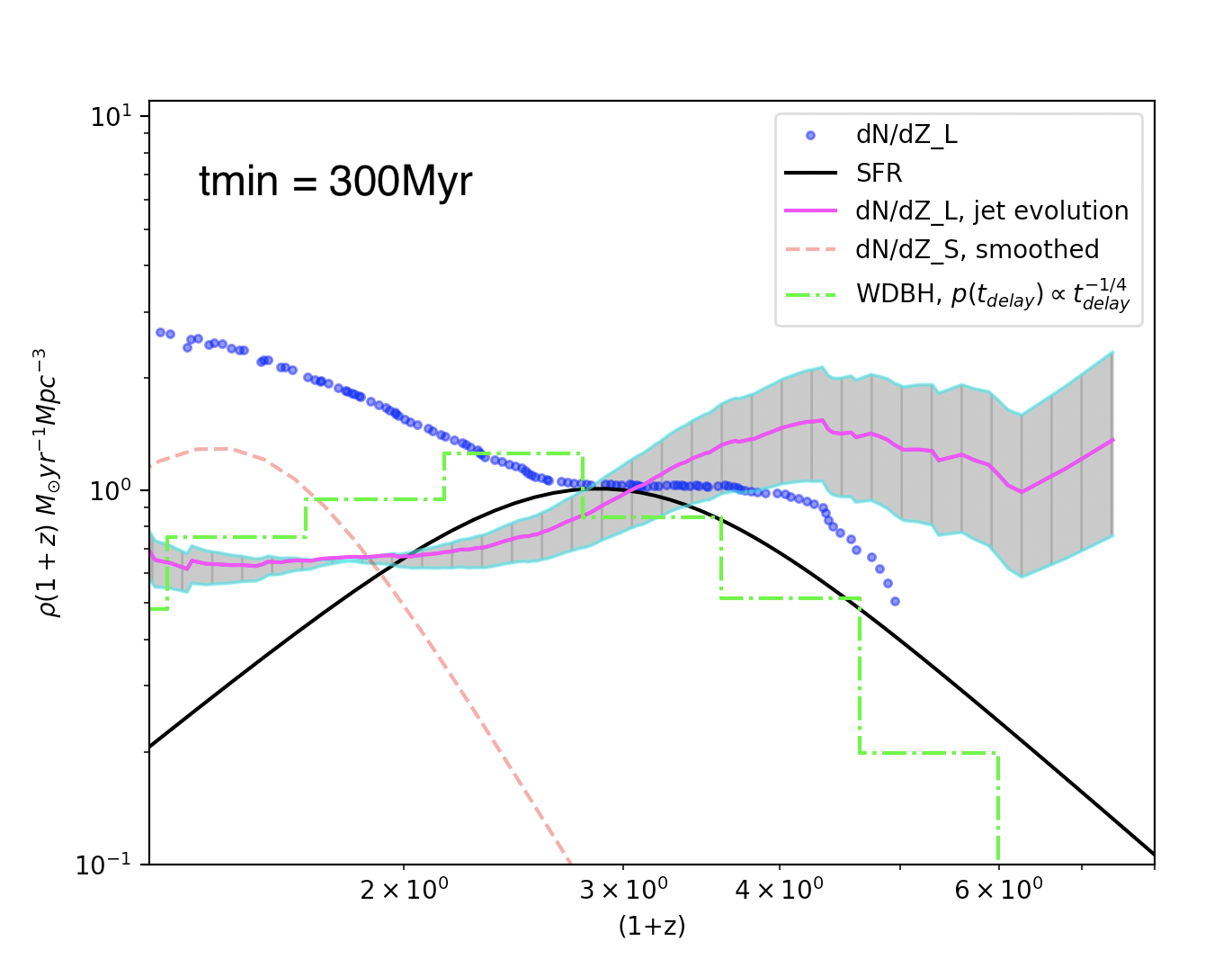}\includegraphics[width=8cm,height=6.25cm]{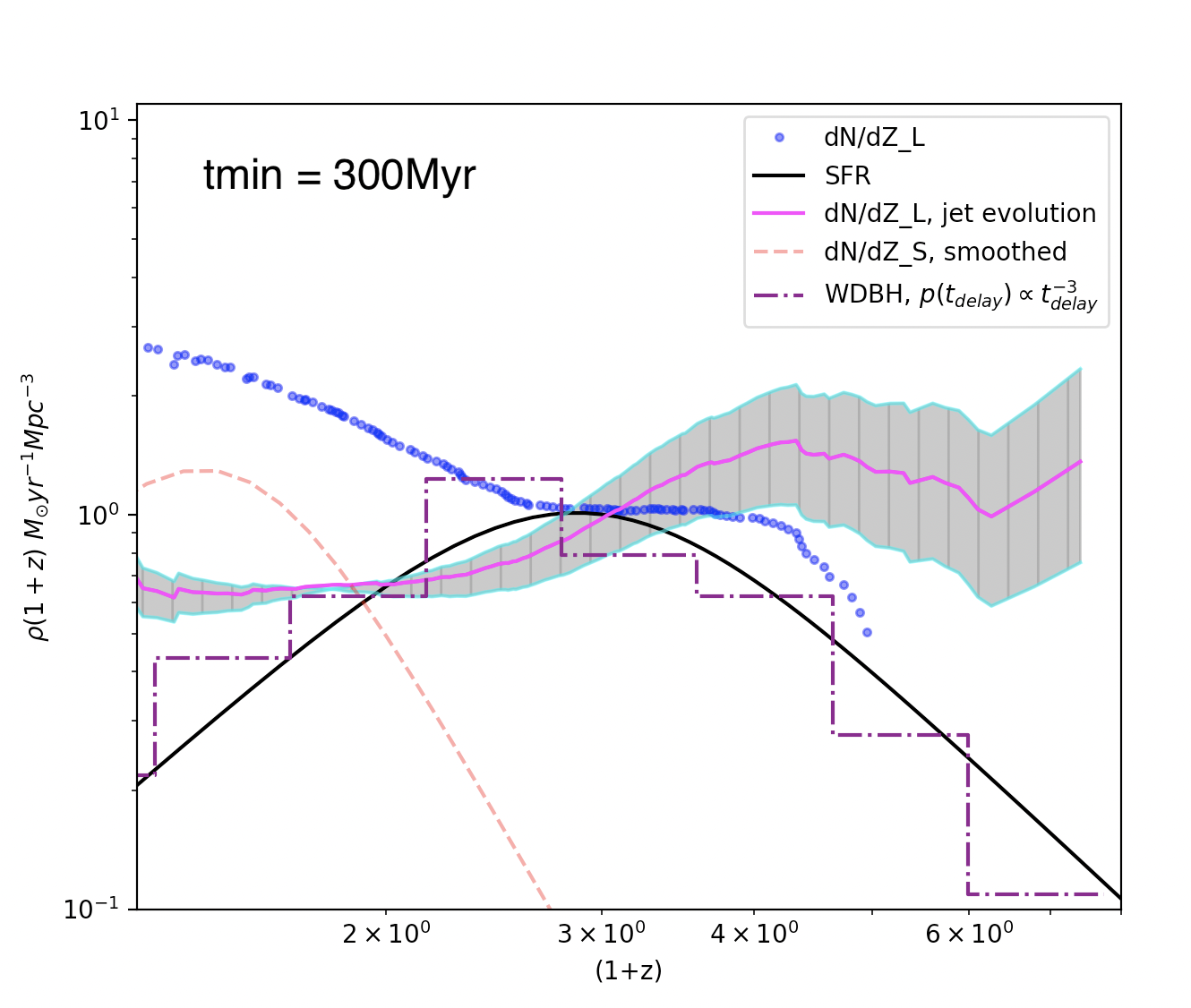} \\
\includegraphics[width=8cm, height=6.25cm]{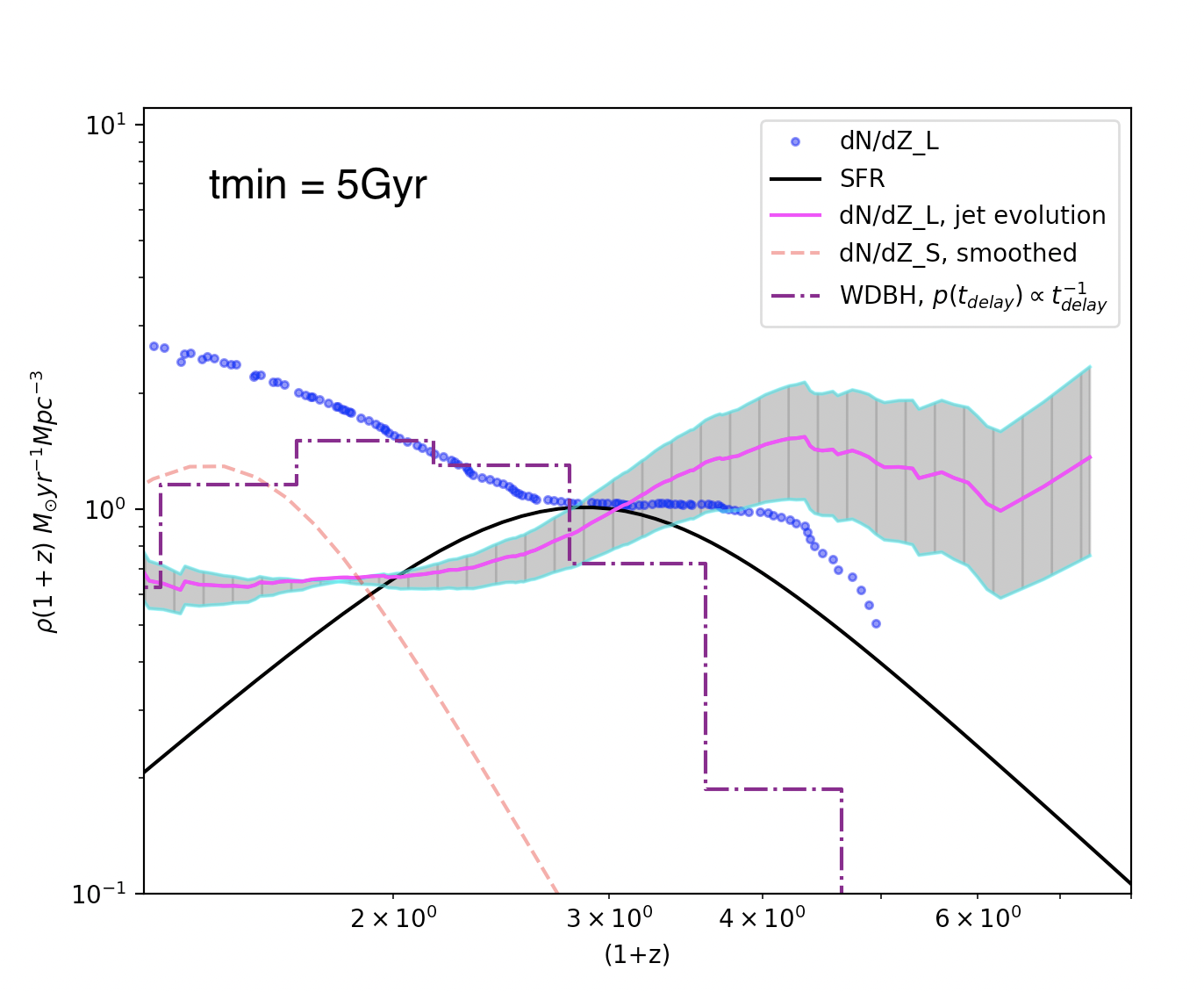}\includegraphics[width=8cm, height=6.25cm]{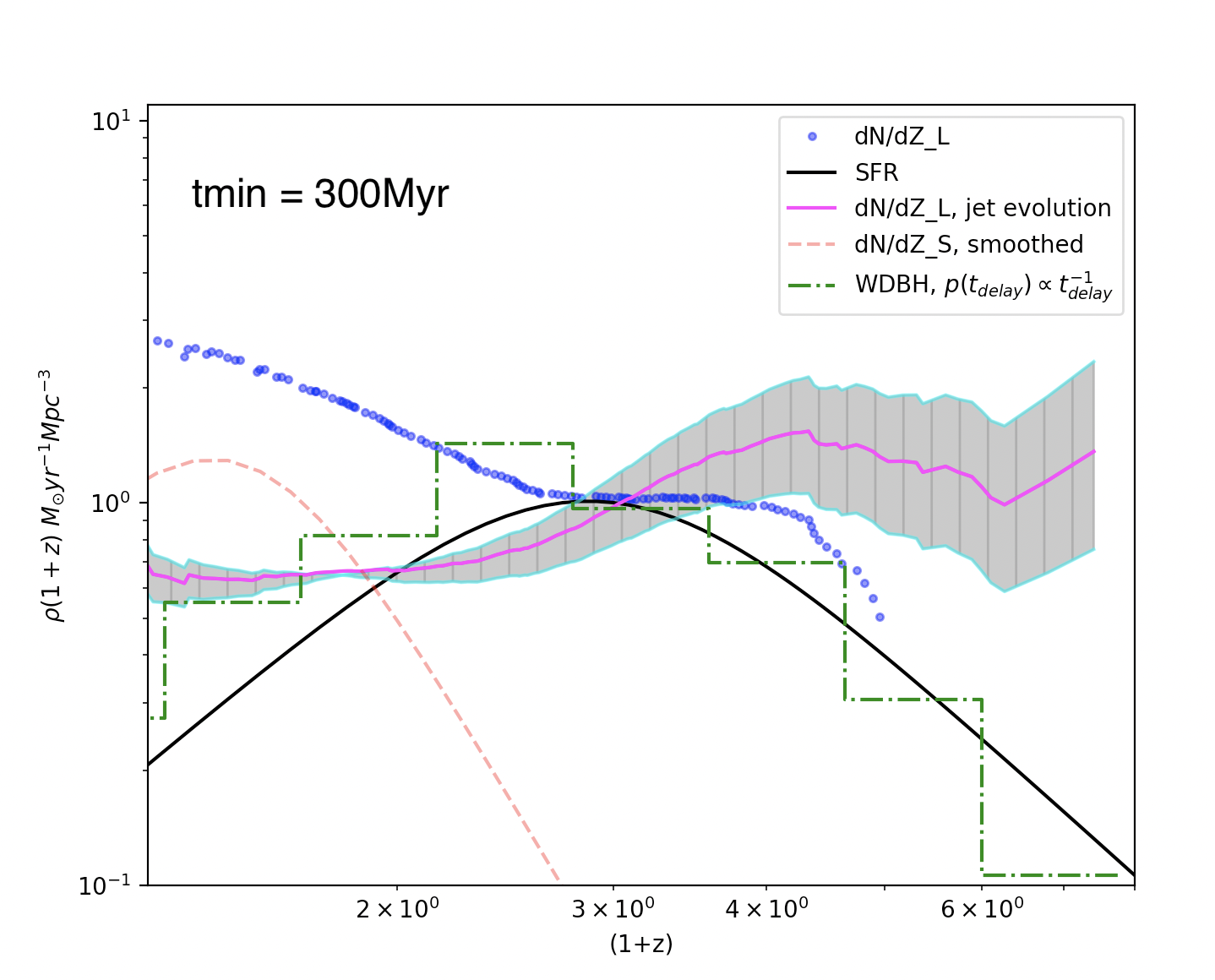}
    \caption{Same as Figure 1, but now we have included our theoretical estimates for the contribution from the WDBH merger population that could produce GRBs (dash-dot histograms). {\bf Top Panels:} The left panel shows our predictions for the WDBH rate density for a very flat DTD with a power law index of $\beta = 1/4$ (see equation 13 for the definition of $\beta$), while the right hand panel shows the results for a steep DTD with a power law index $\beta=3$. Both DTDs have a minimum cutoff time of $t_{min} = 300$ Myr.  It is clear only the flattest/shallowest DTDs can account for a significant uptick, for this value of $t_{min}$. {\bf Bottom Panels:} Both panels show results using a DTD with a power law index of $\beta=1$, but the left hand bottom panel has a minimum cutoff time of $5$ Gyr, compared to $300$ Myr for the right panel.  As expected, a larger $t_{min}$ produces a much greater excess of these systems at low redshift.  }
    \label{fig:example_figure}
\end{figure*}

\begin{figure*}
\includegraphics[width=15cm]{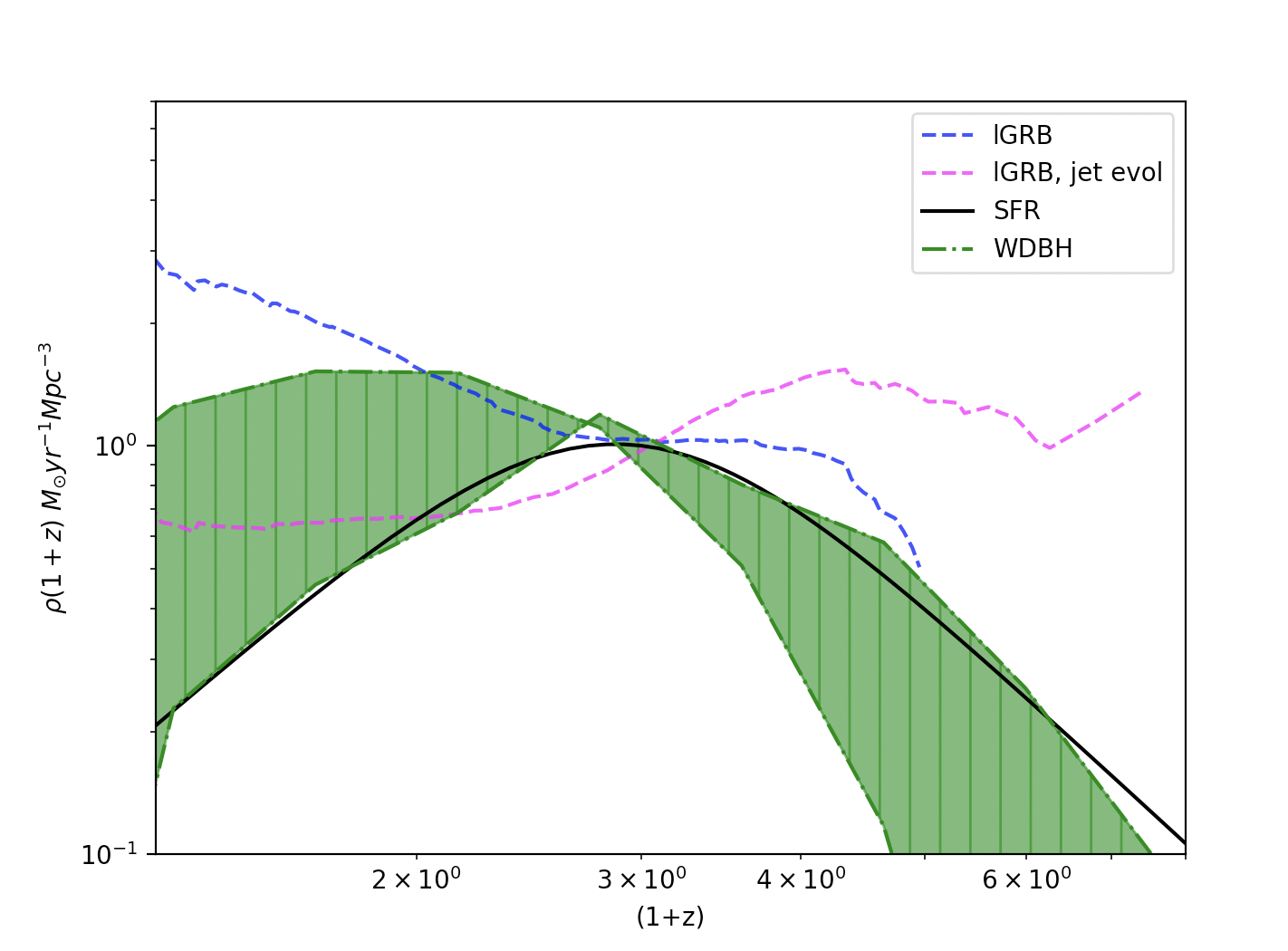} 
    \caption{Same as Figures 1 and 5, but showing only the SFR (black line), the rate density from lGRBs (blue dashed line and magenta line, where the latter corrects for jet opening angle evolution), and a the contribution from WDBH merger systems for a reasonably motivated range of parameter space for the DTDs (green region), as discussed in the text.}
    \label{fig:example_figure}
\end{figure*}



\section{Population Synthesis Simulations and WDBH Rate Density}

We use the open-source code {\em Compact Object Synthesis and Monte Carlo Investigation Code} (COSMIC) to run our population synthesis calculations \citep{Brev20}. This code is specifically tailored for modelling binary systems, computing stellar evolution based on the equations of \cite{HPT00} and binary interactions according to \cite{HPT02}, but includes extensive improvements to account for factors important for compact object formation, such as metallicity-dependent winds and black hole natal kick strength \citep[for a detailed description of the updates see][]{Brev20}.  We note the \cite{HPT00} work develops an extensive set of equations to account for all stages of a star's evolution through the Hertzsprung-Russell diagram, across all masses and metallicities.

The initial sample of binary parameters (primary and secondary masses, orbital periods, eccentricities, metallicities, and star formation histories) is generated by randomly sampling distribution functions based on models described in \cite{MDiS17}. They use a wide range of observational data to develop their models, and find an empirical relationship between mass ratio and orbital separation (with more modest ratios $q \sim 0.5$ for smaller orbital periods, while larger orbital periods seem to have mass ratios consistent with random sampling from the initial mass function), which is incorporated into the COSMIC code.

COSMIC has well-defined prescriptions to deal with binary mass transfer, common envelope evolution, winds, natal kicks, and other binary-specific evolution parameters, based on the equations laid out in \cite{HPT00} and \cite{HPT02}.  We adopt an Eddington-limited mass-transfer scheme, draw natal kicks from a bimodal distribution based on whether they go through an iron core-collapse supernova versus an electron-capture or ultra-stripped supernova, and follow the wind prescription described in \cite{Vink05}. We have also employed a common envelope efficiency (defined as the efficiency of transferring orbital angular momentum into kinetic energy of the envelope) of 1. The onset of unstable mass transfer and common envelope evolution are determined by a critical mass ratio according to \cite{HW87}.

COSMIC takes a Monte Carlo approach to sampling and evolving the population which allows for much faster and more efficient simulations.  For our simulations described below we confirmed convergence, ensuring we sample sufficiently (and that our results do not depend on our computational set-up such as number of processors, threading, etc).\\


We have run a suite of simulations over different initial conditions, focusing on systems in which the initial masses of the stars fall in the range of $1 - 80 M_{\odot}$ and which evolve into WDBH binaries (we note the COSMIC convergence criteria are computed at the formation of the binary).  For a given galaxy we simulate a burst of star formation as well as constant star formation lasting 0.5 Gyr and 10 Gyr, over a range of metallicities.  In Table 1, we provide the simulation output for three representative runs at three different metallicities, showing the fraction of WDBH systems relative to the total number of stars in the simulation.\\

We show the output for the binary separations and the black hole spin from a representative simulation at a metallicity of $Z=10^{-2}$ in the left and right panels of Figures 4, respectively. This simulation was run under the condition of a burst of star formation at a redshift of $z \approx 2$; we then allowed the population of stars evolve from that point to the present day. As mentioned previously, the black hole spin is calculated according to the prescription described in \cite{Bel20}. The y-axis shows un-scaled simulation numbers\footnote{To scale to an astrophysical population, we need to multiply the fraction of our WDBHs by the total number of stars in a given galaxy. For the purposes of this paper, we remain agnostic to this number (i.e. the number of stars in any particular galaxy).}.  Figure 5 shows the corresponding DTD, from the simulation although we caution that this distribution is solely based on loss of angular momentum due to gravitational wave radiation.  \\

Our results indicate that the WDBH fraction per galaxy is $\lesssim 10^{-6}$ of the total number of stars for separations $r \leq 100 R_{\odot}$, across a range of metallicities spanning $Z=10^{-3}$ to $Z=10^{-2}$. Interestingly, we find a slightly higher fraction  of these systems for higher metallicity (although only very marginally so).   This fraction decreases to about $3 \times 10^{-8}$ if we only consider WDBH with $r \lesssim 2 R_{\odot}$ (i.e. at a separation where gravitational radiation is enough to cause the merger within a Hubble time).
We note that \cite{Fry99} estimates the rates of WDBH mergers to be between $10^{-9}$ to $10^{-6}$/yr/galaxy. \cite{Dong18}, using results from \cite{Nel01}, quote a rate of about $2 \times 10^{-6}$/yr in the galactic disk, for systems with a typical black hole mass of about $5 - 7 M_{\odot}$.  \\

In principle we can also estimate an upper limit to this rate using observations of the low mass X-ray binary (LMXB) population.  For example, if we take the average LMXB luminosity and use the scaling relation between total X-ray (in the range of 0.5 - 8 keV) LMXB luminosity per galaxy and stellar mass, $L_{\rm LMXB}/M_{\rm star} \sim 9 \times 10^{28} {\rm erg} {\rm s}^{-1} {\rm M}_{\odot}^{-1}$ \citep{Gilf04, Leh10, US18}, we can roughly estimate the number of LMXBs for a given stellar mass.  Figure 12 of \cite{Grim02} shows the luminosity function of LMXBs, which can vary across several orders of magnitude \cite{Fab06}. Recent observations of LMXBs in our galaxy have cataloged roughly $\sim 300$ LMXBs in the Milky Way \citep[e.g.][]{Avakyan2023,Fort24}. However, this number is difficult to pin down and likely a lower limit because of the transient nature of these systems as well as accounting for detector sensitivity limits, but it can help guide our rate estimates in principle, keeping in mind the WDBH population is only a fraction of the LMXB population.\\




\begin{table}
    \centering
        \caption{Fraction of white dwarf-black hole binaries to total number of stars for different metallicities from the population synthesis code COSMIC}
     \begin{tabular}{cc}
    \hline 
        
        \hline
        \hline
         $n_{WDBH}/n_{stars}$&  Z \\  \hline
         $7. (\pm 1.) x10^{-7}$&  $.001$ \\ 
         $3. (\pm 1.) x10^{-6}$&  $.005$ \\ 
         $1. (\pm 1.) x10^{-6}$&  $.01$ \\ 
\hline
\hline

    \end{tabular}
    \label{tab:angles}
\end{table}


\subsection{WDBH Rate Density}
Ultimately, we are trying to understand if the uptick at low redshifts in the rate density of long GRBs (shown by the cyan, blue, and magenta lines in Figure 1) can be explained by WDBH mergers, under the assumption that the WDBH merger rate density follows the star formation rate {\em convolved with a delay time distribution}, which accounts for the time between formation of the system and merger.  \\

\noindent Our approach to estimating this is as follows:
\begin{enumerate}
    \item For a given so-called birth redshift, $z_{birth}$, of a WDBH system, we draw a sample of 1000 delay times from a delay time distribution as parameterized in equation 13, with a power-law index $\beta$ and a minimum delay time $t_{min}$\footnote{It is important to take care when transforming from time to redshift space; a given merger delay time will correspond to a different redshift delay at different redshifts.}.\\
    
    \item For each delay time, we compute the corresponding redshift at which the merger would occur, $z_{merge}$, the upper limit of the following integral:\\
    \begin{equation}
        t_{d} = -\frac{1}{H_{o}}\int_{z_{birth}}^{z_{merge}}\frac{dz}{(1+z)E(z)}
    \end{equation}
    where $E(z)$ is given by:
    \begin{equation}
        E(z) = \sqrt{\Omega_{m}(1+z)^{3} + \Omega_{\Lambda}}
    \end{equation}
    and where we use an $H_{o}$ of 70 km/s/Mpc, $\Omega_{m} = 0.3$, and $\Omega_{\Lambda} = 0.7$. \\

    \item We now have a distribution of where WDBH mergers will occur for a given birth redshift, for a particular DTD.  We then normalize this distribution by the star formation rate of \cite{MD14} at the birth redshift, $\rho_{SFR}(z_{birth})$. Note that we do not make any corrections for jet opening angle evolution; this is because we have no a priori theoretical or observational constraints on this factor for WDBH merger systems. This is contrast to lGRBs resulting from the collapse of massive stars where it is expected that the dense stellar envelope and cocoon environment around the black hole disk system will collimate the jet as it emerges into the medium beyond the stellar envelope and radiates \citep{LRH20}.\\

    \item We repeat the steps above for a series of birth redshifts ranging from 10 to 0.5, and then integrate all of the the SFR weighted distributions over birth redshifts to get the final total rate density of where the WDBH merger occurs.
    
\end{enumerate}

   Our results are shown in Figures 5 and 6, for a range of potential DTDs for the WDBH systems.  In particular, we allow for a relatively wide range in the power-law index $\beta$ and minimum delay time $t_{min}$. The top panels of Figure 6 show a DTD with a $\beta = 1/4$ (left panel) and $\beta = 3$ (right panel).  As expected, the flatter delay time distributions lead to a larger number of systems at low redshift - an excess compared to the star formation rate at these redshifts.  Both of the bottom panels of Figure 6 show results utilizing a DTD with a $\beta = 1$, but with different minimum delay times.  The left panel, with a 5 Gyr minimum delay time, shows a strong excess in the rate density compared to the star formation rate at low redshifts, while the right panel, with a 300 Myr delay time, shows a smaller excess.\\

   We note that shallow DTDs (i.e. low values of $\beta$) may be well-justified not only because these DTDs are currently very unconstrained, but also because a significant fraction of these systems could form by dynamical capture, as is believed to be the case for LMXBs in globular clusters, for example. This could serve to flatten the DTD and also increase the value of $t_{min}$.  A large value of $t_{min}$ is also motivated simply by the fact that it can take a few Gyr for a lower-mass star to form the WD component of the system.   \\
   
   The green region of Figure 7 highlights the range of parameter space where the WDBH merger systems produce an uptick at low redshifts compared to the star formation rate.  This shaded region covers the range of parameter space for the DTDs shown in Figure 7. The dashed lines are the lGRB rate density with and without correcting for beaming angle evolution (blue and magenta lines, respectively). It appears that under our assumptions, at least some of the low redshift uptick (all, in the case of the beaming-angle-evolution corrected rate density) can be accommodated by WDBH mergers.

\section{Conclusions}

In this paper, we have investigated the viability of a WDBH merger progenitor for long gamma-ray bursts, particularly focusing on its contribution to the apparent low redshift uptick in the GRB rate density.  We have scrutinized how a WDBH merger system can fulfill the necessary observational constraints of lGRBs, specifically the timescales of the prompt emission, the overall energetics, and the rate density. Our main results are as follows:\\

\begin{itemize}
    \item We have provided basic analytic arguments showing that a WDBH merger system fulfills the necessary energy and timescale requirements for an lGRB.  In particular, if most of the mass of the WD is tidally disrupted by its black hole companion and circularized into a disk around a massively spinning black hole, the central engine will be luminous enough and last long enough to power a long GRB.\\
    
    \item We have discussed the many channels of orbital angular momentum loss and connected this to the merger time of the WDBH system, which is ultimately responsible for any deviation of their rate density from the star formation rate. We have shown that this timescale is not well-constrained from first principles, but that constraints from observations of other compact object mergers (e.g. DNS binaries) may be a useful guide when estimating the WDBH delay-time distribution.\\
    
    \item We have run a suite of population synthesis codes across metallicities to estimate the fraction of these systems in a given galaxy, relative to the total stellar population.  We find this fraction ($\sim 10^{-8} - 10^{-6}$) aligns with previously published estimates. We again emphasize we have only focused on systems with WDs that lie on the high end of the mass distribution ($\geq 1 M_{\odot}$).\\
    
    \item  Guided by previously published DTDs for compact object binaries, we estimate the rate density of WDBH mergers and find that for relatively flat DTDs (with a power law index ranging from $\beta \sim 2$ to $\sim 1/4$) and/or a minimum delay time above of a few hundred Myr to a few Gyr, these sytems can explain the apparent low redshift uptick in the lGRB rate density.  Even under conservative assumptions for their DTDs, WDBH mergers can account for the the beaming-angle-evolution-corrected low redshift rate density. The more severe uptick seen in the uncorrected lGRB rate density can be explained by WDBH mergers with shallow DTDs and/or long minimum cutoff times.   

\end{itemize}

We hope to further test this WDBH progenitor model as we obtain more data, in particular examining whether this potential low redshift sub-population of GRBs may be preferentially radio dark, have a unique gamma-ray duration distribution, jet opening angle distribution, and/or larger galaxy offsets on average. Another unique signature may come from quasi-periodic modulations in the light curves, resulting from repeated partial disruptions of these events \citep{Chen24}. Future gravitational wave observations, both with ground-based observatories like LVK \citep{LVK20} and Cosmic Explorer \citep{Ng21}, as well as space-based observatories including LISA \citep{Sberna2021} and higher frequency decihertz graviational wave detectors like DECIGO \citep{Seto01, Kinu22} and TianGO \citep{Kuns20}, may uncover the population of WDBH binaries and directly test their connection to long GRBs as well. \\

 \cite{Church17} has shown that for WDBH binaries to merge, the mass ratio in the binary must be larger than about 0.2, meaning these systems will merge primarily for black hole masses $\lesssim 8 M_{\odot}$.  In subsequent papers \citep{Zen20, Bob22} they suggest that both WDNS and WDBH systems may be accompanied by a faint optical transient, potentially detectable by deep follow-up observations of GRBs. On the other hand, we note an important point when considering this model for GRBs with an observed kilonova signal: \cite{Metz12, Marg16, Rod19} have shown that neither WDNS mergers nor WDBH mergers have sufficient material nor the accretion rate to produce a kilonova signal in the light curve \citep[also noted in][for WDNS mergers]{Bob22}.  In particular, in their calculations, the accretion rate does not exceed a critical value of ($\sim 10^{-2} M_{\odot}/s$) necessary for r-process elements to be produced in the disk-wind ejecta.  As such, we may in fact rule out the WDBH progenitor for those GRBs with a kilonova signal in their light curves. \\

As mentioned in \S 2, some studies \citep[e.g.][]{Per16} have suggested that if a metallicity dependent factor is included in the formation efficiency of lGRBs (where it is assumed that GRBs form more readily in low metallicity environments), the GRB rate density more closely tracks the star formation rate at low redshifts.   Athough theoretical \citep[e.g.][]{HMM05,Yoon06,WH06} and observtional \citep[e.g.][]{GF13, GF17} studies have suggested lGRBs favor low metallicity environments, there remain many open questions regarding the details of the exact metallicity requirements.  Indeed some lGRBs have been observed in high metallcity environments \citep{Lev10, Sav12, Kru12, Ell12, Ell13, H13}, and it is unclear if and how to account for a metallicity cutoff in a definitive and accurate way. \\

Nonetheless, regardless of the presence of the low redshift uptick in the lGRB rate density, we would argue that the study of the contribution of compact object binary progenitors to the lGRB population can help us home in on important factors in the evolution of these systems, as well as better understand the varied GRB progenitor landscape.\\




\section*{Acknowledgements}
 We are very grateful to the anonymous referee for helpful comments that improved this manuscript.  We thank Ayanah Cason, Lailani Kenoly, Katie Breivik, Vera Delfavero, Mike Zevin, Gabriel Casabona, and Shane Larson for helpful discussions related to {\em COSMIC} simulations.  We also thank Brian Metzger and Alexey Bobrick for insightful comments and suggestions.
This work was supported by the U.~S. Department of Energy through Los Alamos National Laboratory (LANL).  LANL is operated by Triad National Security, LLC, for the National Nuclear Security Administration of U.S. Department of Energy (Contract No. 89233218CNA000001).   Research presented was supported by the Laboratory Directed Research and Development program of LANL project number 20230115ER. We acknowledge LANL Institutional Computing HPC Resources under project w23extremex. Additional research presented in this article was supported by the 
Laboratory Directed Research and Development program of Los Alamos National Laboratory under 
project number 20210808PRD1. LA-UR-24-28685

\section*{Data Availability}

 The data generated and used in this paper are available upon request.



\bibliographystyle{mnras}
\bibliography{ms} 








\bsp	
\label{lastpage}
\end{document}